\documentstyle[epsfig,rotating,twocolumn]{mn}

\title[X-ray/Optical Bursts from GS\,1826--24]
{X-ray/Optical Bursts from GS\,1826--24}
\author[A.K.H. Kong, L. Homer, E. Kuulkers, P.A. Charles and A.P. Smale]
{A.K.H. Kong$^1$, L. Homer$^1$, E. Kuulkers$^{2,3}$, P.A. Charles$^1$
and A.P. Smale$^4$\\
$^1$ Department of Astrophysics, Nuclear \& Astrophysics Laboratory,
Keble
Road, Oxford OX1 3RH \\
$^2$ Space Research Organization Netherlands, Sorbonnelaan 2, 3584 CA
Utrecht, The Netherlands \\
$^3$ Astronomical Institute, Utrecht University, P.O. Box 80000, 3507 TA
Utrecht, The Netherlands \\
$^4$ Laboratory for High Energy Astrophysics, Code 660.2, NASA Goddard
Space Flight Center, Greenbelt, MD 20771, USA\\
}

\date{Accepted. Received.}

\begin{document}

\maketitle

\begin{abstract}

We report results from the first simultaneous X-ray ({\it RXTE}) and optical
(SAAO) observations of the low-mass X-ray binary GS\,1826--24 in June 1998.
A type-I burst was detected in both X-ray and optical wavelengths.  Its
energy-dependent profile, energetics and spectral evolution provide evidence for an increase in the X-ray burning area but not for photospheric
radius expansion.  However, we may still derive an upper limit for its  distance of $7.5\pm0.5$ kpc, assuming a peak flux of $\sim 2.8\times 10^{-8}$ 
erg cm$^{-2}$ s$^{-1}$.
A $\sim 3$ s optical delay with respect to the X-ray burst is also
observed and we infer
that this is related to the X-ray reprocessing in the accretion disk into the
optical. This provides support for the recently proposed orbital period of $\sim$2 h. We also
present an {\it ASCA} observation from March 1998, during which two X-ray
bursts were detected. 
\end{abstract}

\begin{keywords}
accretion, accretion disks -- binaries: close -- 
stars: individual (GS\,1826--24) -- X-rays: bursts
\end{keywords}

\section{Introduction}

GS\,1826--24 was discovered serendipitously in 1988 by the
{\it Ginga} satellite (Makino et al. 1988) at an average flux of 26 mCrab 
(1--40 keV) and was fitted by a
single power law spectrum with $\alpha \sim 1.7$. Whilst showing some
evidence for variability during 1988--89 (Tanaka \& Lewin 1995; In't Zand
1992), {\it ROSAT} PSPC observations in 1990 and 1992 (Barret et al.
1995) found comparable
flux levels and no X-ray bursts were detected during 8 hours 
exposure on the source. The spectrum was well fitted by a single power law
with $\alpha \sim 1.5-1.8$ and an absorption column, $N_H \sim 5\times 10^{21}$
cm$^{-2}$.
Temporal analysis of both the {\it Ginga} and {\it ROSAT} data yielded
a featureless $f^{-1}$ power spectrum extending from $10^{-4}$--500 Hz
(Tanaka \& Lewin 1995; Barret et al. 1995), with neither quasi-periodic
oscillations (QPOs) nor pulsations being detected.

Since there was no detection prior to {\it Ginga}, the source was catalogued
as an X-ray transient. Its similarities to Cyg X--1 and GX\,339--4 in the
low state, both in spectrum and temporal behaviour (hard X-ray spectrum and
strong flickering), led to an early suggestion by Tanaka (1989) that it
was a soft X-ray transient with a possible black-hole primary. Following
its detection by {\it CGRO} OSSE in the 60--200 keV energy range,
Strickman
et al. (1996) doubted the suggestion of a black-hole primary after
examining
the combined spectrum from both {\it Ginga} and OSSE. They
found that this required a model with an
exponentially cut-off power law plus reflection term.  The observed cut-off
energy around 58 keV is typical of the cooler neutron star hard X-ray
spectra.  The suggestion that
GS\,1826--24 contains a neutron star was also discussed in detail by
Barret et al. (1996), where they compared the luminosity of the source
with other X-ray bursters. The recent report of 70 X-ray bursts in 2.5
years by {\it BeppoSAX} WFC (Ubertini et al. 1999) and an optical burst
by Homer et al. (1998) confirms the presence of a neutron star accretor. 

Following the first {\it ROSAT} PSPC all-sky survey observations in
September
1990, and the determination of a preliminary X-ray position, a search for
the counterpart yielded a time variable, UV-excess, emission line star
(Motch et al. 1994; Barret et al. 1995). The source had $B = 19.7$, and
an uncertain V magnitude ($\sim 19.3$), due to contamination by
a nearby star. Subsequent high-speed CCD photometry by Homer et al.
(1998) yielded a $\sim$ 2.1 hr optical modulation, but confirmation of
its stability requires observation over a longer time
interval. We therefore carried out an {\it ASCA} observation and
simultaneous {\it RXTE}/optical observations of GS\,1826--24
in order to study its spectral behaviour and very short timescale  
variability, as well as the 2.1 hr optical modulation. In Table 1, we
summarize the {\it ASCA}, {\it RXTE} and SAAO observations used in this
work.

This paper is structured as follows. An outline of all the X-ray and optical
observations is given in section 2. In section 3 we report the
spectral analysis of {\it ASCA} data for both persistent and burst
emission. Simultaneous {\it RXTE}/optical
observations, including the analysis of a simultaneous X-ray/optical
burst are also presented. We discuss the implications of the 
X-ray bursts and constrain the nature of the source from the delay between 
the X-rays and optical in section 4. 
We present the overall timing study of the {\it ASCA} and {\it
RXTE}/optical observations in a companion paper (Homer et al. 1999,
henceforth paper II).

\begin{table}
\caption{A Journal of the X-ray/optical observations of GS\,1826--24} 
\begin{tabular}{c c c c c} \hline
Observatory&Detector& Date&\multicolumn{2}{c}{Time (UT)}\\
&                                          &         &Start &End\\
\hline
{\it ASCA}& SIS+GIS            &31.03.98&      10:52&20:54\\
{\it RXTE}& PCA                &23.06.98&      19:20&21:59\\
&                              &24.06.98&      19:20&21:58\\
&                              &25.06.98&      19:21&21:57\\
&                              &28.07.98&      08:12&14:22\\
&                              &29.07.98&      13:05&15:51\\
SAAO& 1.9m+UCT CCD             &23.06.98&      19:08&04:57\\
&                              &24.06.98&      19:07&03:17\\
&                              &25.06.98&      18:36&04:11\\
\hline
\end{tabular}
\end{table}

\section{Observations and Data Reduction}

\subsection{{\it ASCA}}

The {\it ASCA} satellite consists of four co-aligned telescopes, each of
which is a conical foil mirror that focuses X-rays onto two Solid State
Imaging Spectrometers (SIS) and two Gas Imaging Spectrometers (GIS)
(Tanaka, Inoue \& Holt 1994). The
SIS detectors are sensitive to photons in the 0.4--10.0 keV energy band with
nominal spectral resolution of 2\% at 6 keV. The GIS detectors provide
imaging in the 0.7--10 keV energy range and have a relatively modest
spectral resolution of 8\% at 6 keV, in comparison to the SIS, but with
a larger effective area at higher energies.

For our observation of GS\,1826--24 on 1998 March 31 (see Table 1), one
CCD was activated for each SIS, giving an 11$'$ 
$\times$ 11$'$ field of field and temporal resolution of 4 s. The  
GIS detectors were set to MPC mode (i.e. no image could be extracted) so
that the temporal resolution would be improved to 0.5 s.

The data were filtered with standard criteria including the rejection of 
hot and flickering pixels and event grade selection. We extracted the SIS
source spectra from circular regions of $3'$ radius, yielding 11.2
ks total exposure. The background spectra were extracted from
source-free regions of the instruments during the same observation. For GIS,  
after the standard selection procedure, the net on-source time was 18.9
ks.

\subsection{{\it RXTE}}

We also observed GS\,1826--24 with the Proportional Counter Array (PCA)
instrument on {\it RXTE} (Bradt, Rothschild \& Swank 1993) between 1998
June 23 and July 29 (see Table 1). The PCA consists of five nearly
identical Proportional Counter Units (PCUs) sensitive to X-rays with
energy range
between 2--60 keV and a total effective area of $\sim$ 6500\,cm$^2$. The
PCUs each have a multi-anode xenon-filled volume, with a front propane  
volume which is primarily used for background rejection. For the entire 
PCA and across the complete energy band, the Crab Nebula produces a count
rate of
13,000 counts s$^{-1}$. The PCA spectral resolution at 6 keV is
approximately
18\% and the maximum timing resolution available was 1 $\mu$s. However, in
order to maximize
our timing and spectral resolution, we adopted a 125$\mu$s time
resolution, 64 spectral
energy channel mode over 2--60 keV in addition to the standard mode
configuration. All light curves and spectra presented here have been
corrected for background and dead-time. For most of the time, at   
least four PCUs were turned on, and so we utilized only data from these  
PCUs in order to minimize systematic uncertainties.

\subsection{Optical}

Observations of a small ($50\times 33$ arcsecs) region surrounding the
optical
counterpart of GS\,1826--24 were made using the UCT-CCD fast photometer
(O'Donoghue 1995), at the Cassegrain focus of the 1.9m telescope at SAAO,
Sutherland from 1998 June 23 to 26. The UCT-CCD fast photometer is a   
Wright Camera 576$\times$420 coated GEC CCD which was used here
half-masked so as
to operate in frame transfer mode, allowing exposures of as short as 2~s
with no dead-time. The conditions were generally good with typical seeing
$\sim$ 1.5--2.5 arcsec and the timing resolution for the source was 5 s.
An observing log is presented in Table 1. We
performed data reduction using IRAF, including photometry with the 
implementation of DAOPHOT II (Stetson 1987). Due to moderate crowding
of the counterpart with a nearby but fainter neighbour and the variable  
seeing, point spread function (PSF) fitting was employed in order to obtain
good photometry.  The details of this procedure are given in Homer et al.
(1998).

\section{Analysis and Results}

\subsection{{\it ASCA} Observations}

Given the better spectral resolution and higher sensitivity below 2 keV
of the SIS detectors, we use those data to study
the persistent emission of GS\,1826--24. The spectrum
(excluding the burst intervals, see below) was fitted with a blackbody
plus a power law component . The fit
quality was good, with $\chi^2_{\nu}=1.14$ for 307 degrees of freedom 
(d.o.f.) and these results are summarized in Table 2 and Fig. 1 (whenever
an error for a spectral parameter is quoted throughout this paper, it
refers to the single parameter 1$\sigma$ error).

\begin{figure}
{\rotatebox{-90}{\epsfxsize=\columnwidth\psfig{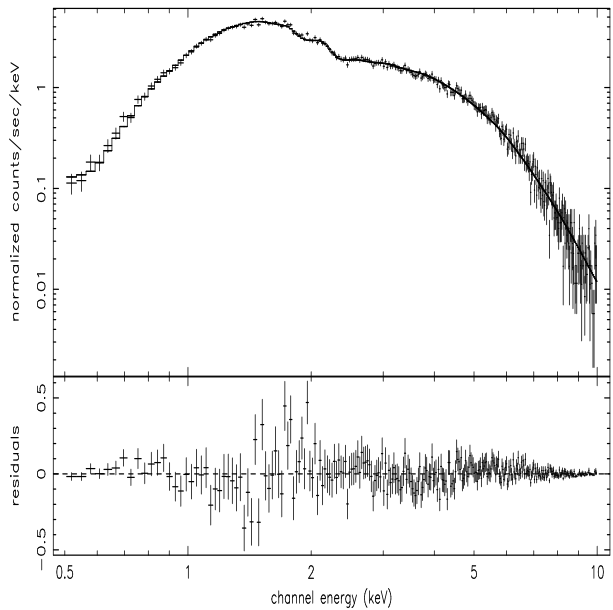}}}
\caption{Upper panel: {\it ASCA} SIS spectral fit to GS\,1826--24
persistent emission. The spectrum was
fitted with a blackbody ($kT=0.74\pm0.02$ keV) plus a power law
component ($\alpha=1.11\pm0.09$). Lower panel: residuals in units of
$\sigma$.}
\end{figure}

\begin{table}
\begin{minipage}{70mm}
\caption{GS\,1826--24 persistent emission SIS spectral fit. The errors
are single parameter 1$\sigma$ errors.}
\begin{tabular}{c c}
\hline
Model& Power law plus blackbody \\
\hline
$N_H$& $(4.0\pm0.36) \times 10^{21}$ cm$^{-2}$ \\
$kT_{bb}$\footnote{Blackbody temperature}& $0.74\pm0.02$ keV \\
$R_{bb}$ at 8 kpc \footnote{Blackbody radius}& $11.3\pm0.6$
km \\
Photon index, $\alpha$& $1.11\pm0.09$ \\
$\chi^2_{\nu}$ (307 d.o.f.)& 1.14 \\
Flux 0.5--10 keV& $(6.8\pm0.2) \times 10^{-10}$ erg cm$^{-2}$ s$^{-1}$
\\
Flux 2--10 keV & $(5.9\pm0.2) \times 10^{-10}$ erg cm$^{-2}$ s$^{-1}$
\\
\hline
\end{tabular}
\end{minipage}
\end{table}

Our results are consistent with those obtained by In't Zand et al.
(1999) using the {\it BeppoSAX} NFI, taken six days later.
Moreover, their 2--10 keV flux of $5.41 
\times 10^{-10}$ erg cm$^{-2}$ s$^{-1}$ which is only $\sim 9\%$ smaller 
than our determination, is consistent with the $\sim 10\%$ fall in count
rate seen by the {\it RXTE} ASM during that interval.

Two X-ray bursts were detected by the {\it ASCA} GIS and one of them was
caught by SIS. The time interval between the two bursts was $\sim$ 5.4
hr and is consistent with the $5.76\pm0.6$ hr quasi-periodicity of the
burst recurrence
as found by {\it BeppoSAX} WFC observations (Ubertini et al. 1999).
Figure 2 shows the time profiles of the two bursts in the 0.7--10 keV
range with 4 s binning. Their rise times (difference between time of the
peak and time of the start of the burst using a linear-rise
exponential-decay model) and e-folding times are comparable, see Table 3.

\begin{figure}
{\rotatebox{-90}{\psfig{file=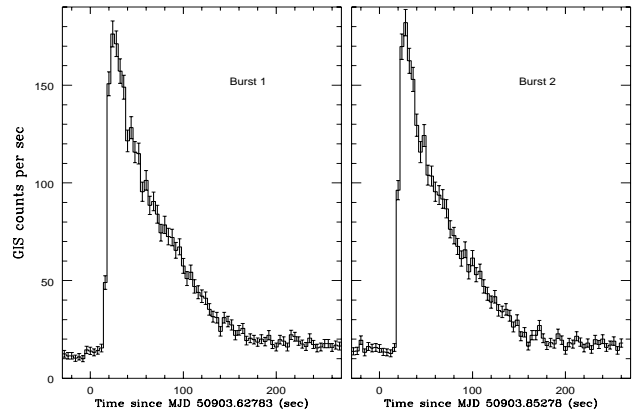,height=9cm,width=6cm}}}
\caption{The time profiles of the two X-ray bursts in the 0.7--10 keV 
range at a time resolution of 4 s as observed by the {\it ASCA} GIS in   
March, 1998.}
\end{figure}

\begin{table}
\caption{Timing parameters of the two bursts by fitting linear-rise
exponential-decay model.}
\begin{tabular}{c c c}
\hline
& burst 1 & burst 2\\
\hline
Peak time (MJD) & 50903.628 & 50903.853\\
Rise time (s) & $6.7\pm0.2$ & $8.8\pm0.02$ \\
e-folding time (s) & $55\pm1.8$ & $50.8\pm1.9$\\
Peak flux (0.7--10 keV; Crab units) & 0.22 & 0.23 \\
\hline
\end{tabular}
\end{table}

Since the GIS was set to MPC mode, which has no positional
information, the lack of background estimation limits the usefulness 
of the spectra. We therefore here analyse the first burst which was
detected with SIS.

We extracted a series of spectral slices through the burst with 4 s time
resolution in the rising phase and 20 s resolution during decay.
Spectral analyses of these slices were performed over the
0.5--10 keV range using a variety of approaches. The most straightforward
`standard' approach was to choose a 300 s section of data immediately
prior to the burst and use this as our `background' for spectral fits to
the individual spectra through the burst. The net (burst --
`background') emission was well fitted ($\chi^2_{\nu} \sim
0.6-1.2$) with a simple blackbody. Figure
3 (left) shows the time variation of the bolometric flux, blackbody
temperature and radius assuming a distance of 8 kpc (In't Zand et al.
1999). The radii are rather low ($\sim 4$ km) and show an
anti-correlation with temperature.

However, the analysis of X-ray burst data can be complicated in cases
where the persistent emission contains a blackbody contribution from the
outer layers of the neutron star. Failing to account for this component
may lead to errors in the temperature determination and severe
underestimates in
the derived blackbody radius during the later stages of the burst (van
Paradijs \& Lewin 1985). We thus repeated our analysis, fitting the above
two-component model to each gross (continuum + burst) spectrum, rather   
than a single blackbody component to the net burst spectrum. The  
power-law component was held constant at its continuum emission value
while the blackbody component was permitted to vary. The results of our
two-component spectral fits are shown in Figure 3 (right), once again for
an assumed distance of 8 kpc. The most important   
difference with the results of the `standard' approach is that the
blackbody radius is now in the range for a typical neutron star ($\sim$
10 km). The blackbody radii show moderate variations, which appear to be
anti-correlated with temperature. This indicates that the blackbody
radiation from the neutron star contributes significantly to the
persistent emission. The blackbody temperatures are higher during the
beginning of the burst and then decline as expected for a type I burst (see
Figure 3). 

\begin{figure*}
{\rotatebox{-90}{\psfig{file=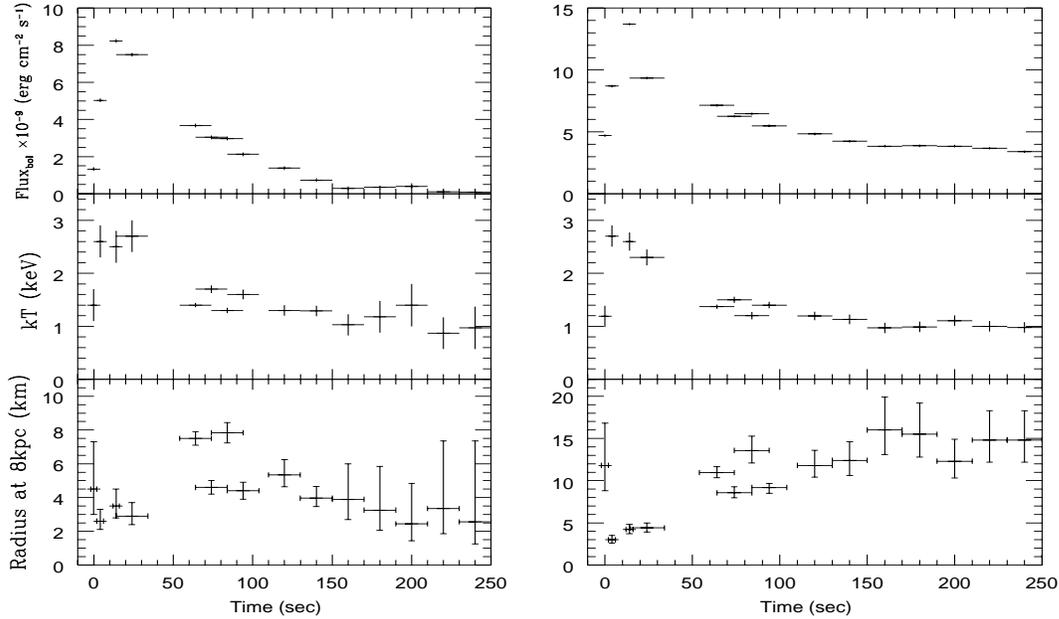,height=15cm,width=9cm}}}
\caption{Results of the spectral analysis of net burst (left) and gross
burst (right) emission from {\it ASCA}. Plotted are the variation of the
flux (top), blackbody temperature (middle) and blackbody radius
(bottom).}
\end{figure*}

The change of the apparent blackbody radius may be affected by the
non-Planckian shape of the spectrum of a hot neutron star (see Sztajno et
al. 1986 and references therein). As a result, the temperature fitted using
a blackbody is simply a `colour temperature' ($T_{bb}$) which is higher
than the
effective temperature ($T_{eff}$); the ratio of $T_{bb}/T_{eff}$ increases
with $T_{eff}$ (e.g. London, Taam \& Howard 1984). Moreover the fitted
blackbody radii are also affected, and so we used the average
relation between blackbody radius and temperature obtained for 4U\,1636--53
(Sztajno et al. 1985) to make an empirical correction to the radii
obtained from the above two-component fits (see van Paradijs et al.
1986). As
this method is strictly empirical, it is independent of possible
uncertainties in model-atmosphere calculations (e.g. Sztajno et al. 1986).
We have assumed that the radii are unaffected for $kT_{bb}$$<$1.25 keV
whilst for $kT_{bb}$$>$1.25 keV the radii decrease linearly with temperature
(e.g. van Paradijs et al. 1986).

The blackbody radii obtained following this final stage of the analysis (see
Figure 4) do not show significant differences compared with the gross
spectral analysis except that the radii during the
peak of the burst are now in the range for a typical neutron star. There is
no evidence for photospheric expansion since the radius remains
almost constant throughout the burst (Figure 4).  We also plotted the flux
$F_{bol}$ versus $F_{bol}^{1/4}/kT_{bb}$ but we do not find evidence
for any increase of the X-ray emitting area (Strohmayer, Zhang \&
Swank 1997).  This ratio is a constant proportional to $(R/d)^{1/2}$, where
$R$ and $d$ are the radius and distance of the source if we assume it is
blackbody emission from a spherical surface. The
unabsorbed bolometric peak flux of the blackbody radiation and other burst
parameters are listed in Table 4. The ratio of the average luminosity
emitted in the persistent emission (since the previous burst) and that
emitted in the burst, ${L}_{pers}/L_{burst}=55\pm5$ assuming the separation of
the two bursts is 5.4 hr.

\begin{figure}
{\rotatebox{-90}{\psfig{file=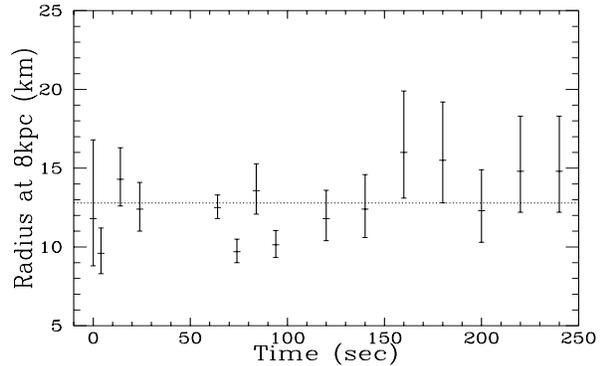,height=8cm,width=5cm}}}
\caption{Variation of the blackbody radius of the gross burst emission
(burst plus persistent) after correction for the deviation of the
spectrum
of a hot neutron star from a pure blackbody spectrum (see text for
explanation). The dashed line is the mean blackbody radius.}
\end{figure}

\subsection{Simultaneous {\it RXTE}/Optical Burst Analysis}

During simultaneous X-ray/optical observations on 24 June 1998, both   
{\it RXTE} PCA and the SAAO 1.9m + UCT CCD detected a burst (see Figure
5). The burst lasted for $\sim$ 150 s and the time profiles of the burst
in both X-ray and optical
are of the fast-rise exponential-decay form with the e-folding
times in the different energy bands given in Table 5. The optical and
X-ray bursts started
almost at the same time but a delay is present between the peaks. The
optical burst resembles the low energy (2--3.5 and 3.5--6.4 keV) X-ray light
curves in which they all have a flat peak and a shoulder during the decay
phase. At higher energies ($>$ 6.4 keV), the peak is much sharper
and the decay is faster during the initial decay phase.

\begin{figure*}
{\rotatebox{-90}{\psfig{file=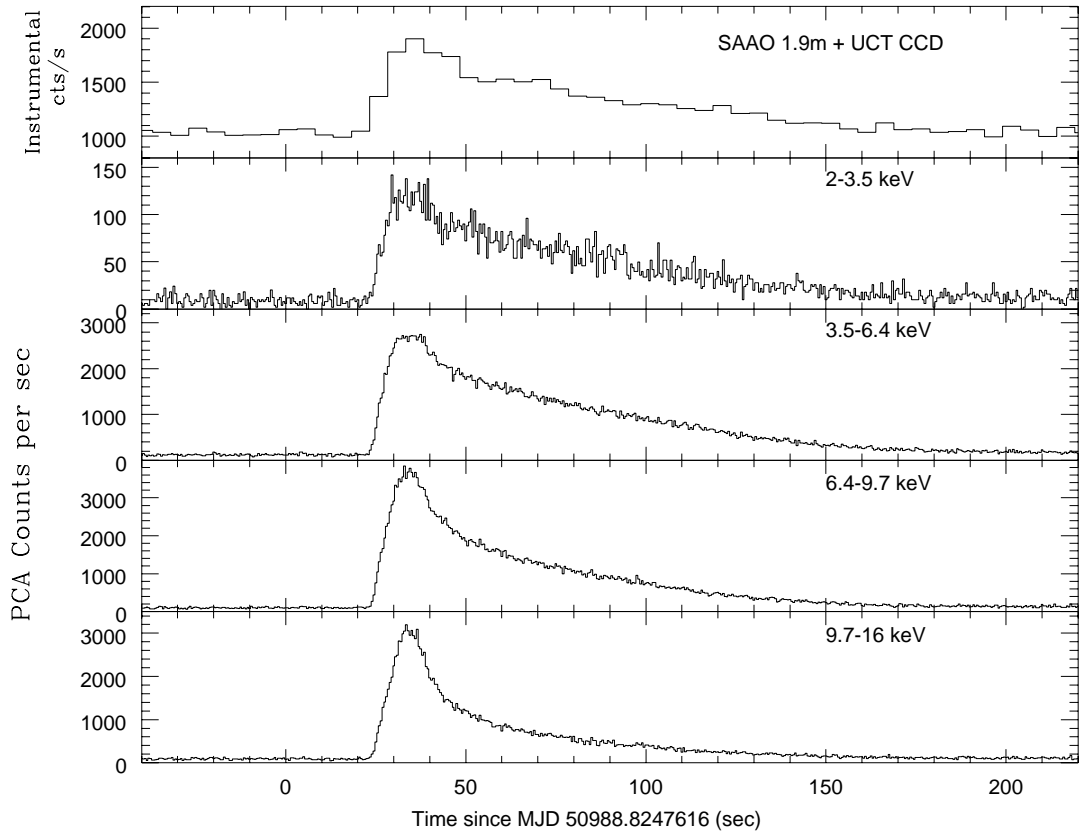,height=16cm,width=12cm}}}
\caption{The optical (SAAO) and X-ray ({\it RXTE}) burst profiles in
various energy bands. The timing resolution is 5 s (optical) and 0.5 s
({\it RXTE}/PCA). The decay times strongly depend on photon energy 
with decays being shorter at higher energies.}
\end{figure*}

\scriptsize
\begin{table*}
\begin{minipage}{150mm}
\caption{Burst parameters for the {\it ASCA} and {\it RXTE} bursts}
\scriptsize{
\begin{tabular}{c c c c c c c c}
\hline
Burst& $F_{max}$ \footnote{Unabsorbed bolometric peak flux} &
$E_{burst}$ \footnote{Total burst fluence}& $E_{pers}$ \footnote{Persistent
emission fluence}&$L_{pers}/L_{burst}$ &
$\gamma$&
       $\tau$&\\
     & ($10^{-8}$ erg cm$^{-2}$ s$^{-1}$)& ($10^{-5}$ erg cm$^{-2}$)
     &($10^{-5}$ erg cm$^{-2}$)& ($E_{pers}/E_{burst}$)&
($F_{pers}/F_{max}$)& ($E_{burst}/F_{max}$)\\
\hline
{\it ASCA}& 3.0&0.13 &7.1 \footnote{Assuming 5.4 hr burst separation}
&54.6& 0.12 & 43.3 s\\
{\it RXTE}& 2.77&0.11&5.5 \footnote{Assuming 5.76 hr burst separation}&
50& 0.1 & 39.7 s\\
\hline
\end{tabular}
}
\end{minipage}
\end{table*}
\normalsize

X-ray spectral analysis is performed in the same way as for the
{\it ASCA} data. The persistent emission is fitted with a single
power-law spectrum with $N_H=(7.3\pm1.5)\times10^{21}$ cm$^{-2}$ and
photon index, $\alpha=1.7\pm0.01$ ($\chi^2_{\nu}=1.11$ for 23 d.o.f.)
which reveals an absorbed flux of $(1.2\pm0.01)\times10^{-9}$ erg cm$^{-2}$
s$^{-1}$ in the 2--20
keV range. The photon index, $\alpha$ is much higher than that seen by {\it ASCA} and suggests a softer spectrum during the {\it RXTE} observations. We
also perform spectral fitting with a power-law plus blackbody model but the fit
does not improve and leads to a large error in $N_H$. Both the 
`standard' method (net burst spectrum) and
gross spectrum were almost indistinguishable from those presented in
Figure 6, presumably because the blackbody provides such a small
contribution to the continuum emission. We also undertook the
non-Planckian analysis as mentioned in the previous section, with results very similar to Figure 6, indicating that the effect due to the
non-Planckian shape of the neutron star spectrum is very small. Once again the neutron star shows the spectral cooling during the burst typical of a type-I
burst.  The
unabsorbed bolometric peak flux of the blackbody radiation and other
burst parameters are listed in Table 4. The ratio
${L}_{pers}/L_{burst}=50\pm4$ if we assume that the separation of two
bursts is 5.76 hr (Ubertini et al. 1999).

The blackbody radius increases to a maximum as the burst
rises, but does not show the simultaneous drop in $kT_{bb}$
and increase in $R_{bb}$ that is the overt signature of photospheric
radius expansion (see Lewin, van Paradijs \& Taam 1995). However, when
we plot the flux $F_{bol}$ versus $F^{1/4}_{bol}/kT_{bb} \propto 
(R/d)^{1/2}$
we do find evidence for an increase in the
X-ray burning area on the star (Strohmayer, Zhang \& Swank 1997). 
This is shown in Figure 7 where the
burst begins in the lower left and evolves diagonally to the upper right and
then across to the left at an essentially constant value until near the end
of the burst. This is an indication that indeed the X-ray burning area
is not a constant but increases with time during the rising phase.
$F^{1/4}_{bol}/kT_{bb}$ eventually reaches a nearly constant value and
the neutron star surface simply cools during the decay phase.

\begin{table*}
\caption{Timing parameters of the simultaneous X-ray/optical burst by
fitting linear-rise exponential-decay model.}
\begin{tabular}{c c c c c c}
\hline
& 2--3.5 keV& 3.5--6.4 keV& 6.4--9.7 keV& 9.7--16 keV& Optical\\
\hline
Peak time (MJD) & 50988.82510& 50988.82511& 50988.82512& 50988.82512&
50988.82516\\
Rise time (s) & $7.2\pm0.2$ & $7.3\pm0.2$& $7.81\pm0.07$& $8.52\pm0.05$&
$10.7\pm0.1$\\
e-folding time (s) & $59\pm2$ & $54.3\pm0.9$& $35.6\pm0.8$&
$20.5\pm0.7$&$55\pm4$\\
\hline
\end{tabular}
\end{table*}

\begin{figure}
{\rotatebox{-90}{\psfig{file=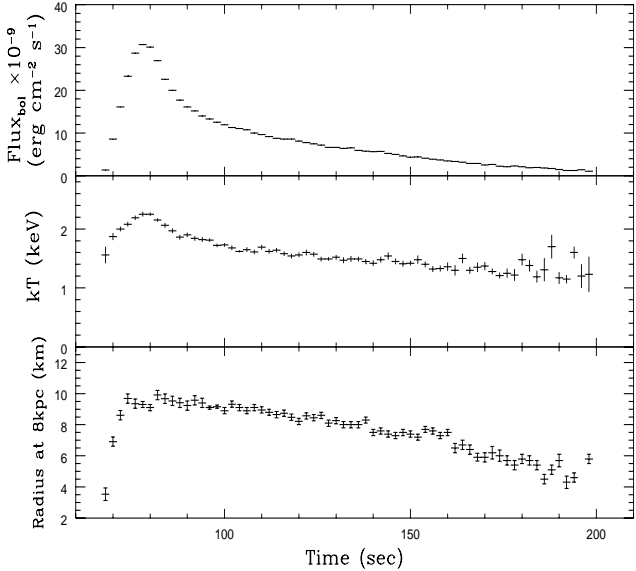,height=9cm,width=8cm}}}
\caption{The variation of the flux, blackbody temperature and
radius during the burst detected by {\it RXTE}. See text for
explanation.}
\end{figure}

\begin{figure}
{\rotatebox{-90}{\psfig{file=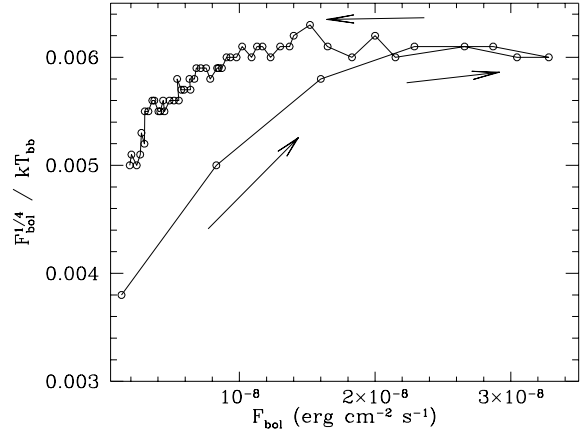,height=8cm,width=6cm}}}
\caption{Plot of bolometric flux $F_{bol}$ vs.
$F_{bol}^{1/4}/kT_{bb}$ for
the {\it RXTE} burst. The burst evolves from the lower left to upper
right and then crosses to the left at nearly constant
$F_{bol}^{1/4}/kT_{bb}$. It is evidence for an
increasing X-ray burning area during the burst rise while the stellar
surface cools off during the decay stage, see text.}
\end{figure}

Kilohertz QPOs between 200 and 1200 Hz were not
detected during the burst with an upper limit of 1\% (at 99\%
confidence).  We also set upper limits on the presence of any  
coherent pulsations during the
burst: $<$ 1\% between 100--500 Hz and 600--1200 Hz, and $<3$\% between
1000--4000 Hz (the Nyquist limit).  A detailed timing analysis of the
remaining simultaneous X-ray/optical data will be presented in paper II.

\subsubsection{X-ray/optical time delay}

Figure 5 shows the simultaneous optical/X-ray burst in different energy 
bands where there is a few seconds delay at the peak of the burst.
In order to quantify this delay, we performed: (i) cross-correlation
analysis, (ii) modelled the optical burst by convolving the X-ray light
curve with a Gaussian transfer function.
\vspace{3mm}

\noindent
{\it 3.2.1.1 Cross-correlation}

We cross-correlated the optical data with X-ray data from different
energy bands as well as the total (2--60 keV) X-ray light curve.
This allows us to determine the correlation and estimate any time lag
between X-ray and optical
variability. The measurement of the cross-correlation function provides
a characteristic delay which does not depend on particular model
fitting. The results show that the optical burst lags the X-ray
burst by $\sim$ 4 s, which is marginally larger than expected in this system.
The separation of the compact object and companion star is 2--3 light-s
if we assume an orbital period of $\sim$ 2.1 hr (Homer et al. 1998), a
neutron star mass of 1.4 $M_{\odot}$ and a companion star mass of
0.1--1.1 $M_{\odot}$.
However, in using a cross-correlation method we are essentially limited
to the 5 s
time resolution of the optical data (the PCA data have 
a much higher time resolution). Moreover, the delay
appears to vary, from almost nothing at the start of the burst to
a few seconds at the peak. This suggests that the delay
might be a function of flux. Therefore a cross correlation analysis cannot
provide a full
picture of the delay between the X-ray and optical fluxes and instead we
model the optical burst by convolving the X-ray light curve with a
transfer function.
\vspace{3mm}

\noindent
{\it 3.2.1.2 Transfer function}

In order to model the time delay between the optical and X-ray bursts, we
convolve a Gaussian transfer function with the X-ray light curve and use 
$\chi^2$ fitting to model the optical light curve. The same method
was used by Hynes et al. (1998) to model the {\it HST} light curve of 
GRO\,J1655--40 from the {\it RXTE} light curve.  The
Gaussian transfer function is given by:

\begin{equation}
\psi(\tau)=\frac{\Psi}{\sqrt{2\pi\Delta\tau}}e^
{-\frac12(\frac{\tau-\tau_0}{\Delta\tau})^2}
\end{equation}

where $\tau_0$ is the mean time delay and $\Delta\tau$ is the dispersion or
`smearing' which is a measure of the width of the Gaussian.
$\Psi$ is the strength of the response.

We performed a series of convolutions of the transfer function with the
lightcurves from the four energy
bands, varying both  $\tau_0$ and $\Delta\tau$ independently.
Essentially
we adjust the overall delay and the degree of `smearing', until
the transferred X-ray lightcurve reproduces the optical response.
Finally,
Figure 8 shows the best fit predicted light curves from each convolution
superimposed on the optical light curve. The principal features of the
optical burst profile are reproduced well in the predicted light curves
from the 2--3.5 and 3.5--6.4 keV energy bands. Table
6 summarizes the results of the Gaussian transfer function fitting to 
all four energy bands. The fits are good ($\chi^2_{\nu}$ $<$ 1 for 51
d.o.f.) for the
two lower energy bands but not for the higher energy ones. The mean
delay and dispersion between the optical and lower X-ray energy bands
are both $\sim$ 3 s (Table 6 and Figure 9). The strength of the
response, $\Psi$, is almost the same for the two lower X-ray energy
bands, at a value
of $\sim 0.0137$, while it is 2--3 times smaller for the higher X-ray
energy band. This is an indication that a greater proportion of the
reprocessing occurs at lower energies.

\begin{figure}
{\rotatebox{-90}{\psfig{file=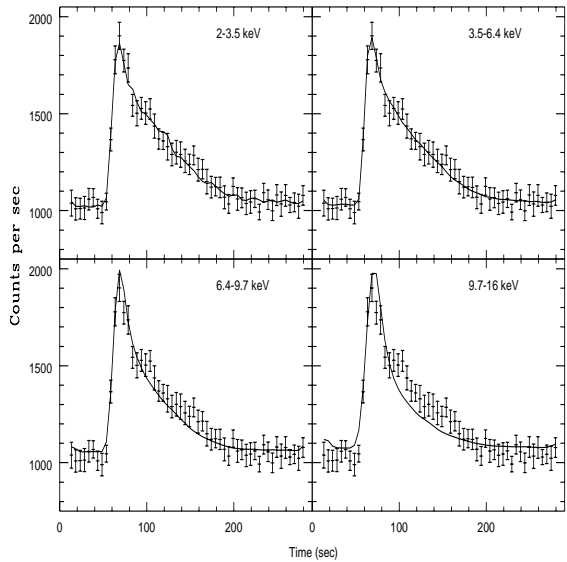,height=8cm,width=8cm}}}
\caption{Best-fitting predicted light curves using a Gaussian transfer function on the four X-ray
energy bands. The resulting curves are superimposed on the optical data points.}
\end{figure}

\begin{table*}
\caption{Summary of results from convolution of a Gaussian transfer
function to the four different X-ray energy bands light curves}
\begin{tabular}{c c c c c c}
\hline
& 2--3.5 keV& 3.5--6.4 keV& 6.4--9.7 keV& 9.7--16 keV\\
\hline
$\tau_0$ (s)& $2.8\pm0.7$& $2.8\pm0.6$& $3.7\pm0.7$& $5.6\pm0.9$\\
$\Delta\tau$ (s)& $3.1\pm1.3$& $2.7^{+1.2}_{-1.7}$& $4\pm1$&
$6.1^{+1.2}_{-1.0}$\\   
$\Psi (10^{-3})$& $13.8\pm6.2$& $13.7^{+2.3}_{-4.1}$& $7.0\pm1.2$&
$4.5^{+0.7}_{-0.5}$\\
$\chi^2_{\nu}$ (51 d.o.f.)& 0.49& 0.42& 1.5& 2.1\\
\hline
\end{tabular}
\end{table*}

\begin{figure}
{\rotatebox{0}{\psfig{file=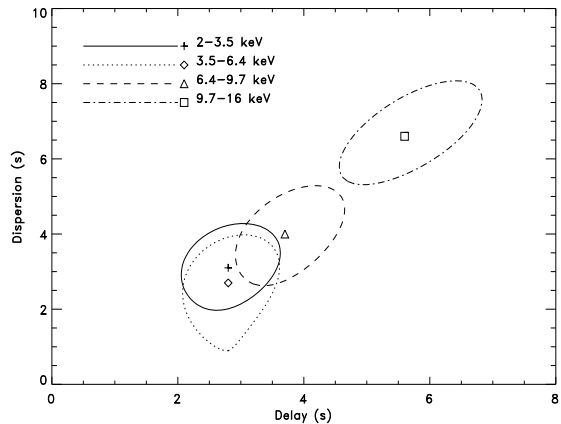,height=6cm,width=8cm}}}   
\caption{Contour plot of the convolution with the {\it RXTE}
light curves and Gaussian transfer function where the contour is defined
as the 1$\sigma$ confidence region.}
\end{figure}

\section{Discussion}

X-ray and optical bursts from GS\,1826--24 were previously reported by
Ubertini et al.
(1999), In't Zand et al. (1999) and Homer et al. (1998).  Our
X-ray/optical observations of GS\,1826--24 showed three X-ray
bursts, of which two were observed by {\it ASCA}, whereas one was 
observed by {\it RXTE} simultaneously with the optical.  On the
basis of their spectral properties
and time profiles, all the
bursts have a cooling trend during their decays and exhibit
blackbody spectra with temperatures of a few keV. The time
profiles show fast rise times of 7--9 s
and long decay times ranging from 20--60 s depending on the
energy band.  We therefore interpret the
three bursts detected from GS\,1826--24 as type I
bursts (Hoffmann et al. 1978). The relatively
long rise time ($\sim 7-9$ s), as compared to bursts in other systems,
indicates that the burst front may have enough time to spread over the
whole neutron star surface during the rise to burst and suggests that the
burning is homogeneous over the surface of the neutron star.  This is
consistent with the fact that we see no evidence for pulsations 
during the burst. We note that X-ray burst rise times in other sources
have been observed to be smaller than $\sim$6 sec (see e.g. Sztajno et al.
1986; Lewin et al. 1987), making this system rather unique. The burst rise
times derived by In't Zand et al. (1999) range from 5--8 s which are also
consistent with our observations.

For the simultaneous burst we may compare the ratio of the persistent to peak fluxes, in both the X-ray and optical, using the results of Lawrence et al.
(1983). They derived a simple power law relation between the changes in
U, B and V band fluxes and the corresponding X-ray flux variations
during a well-studied burst of X1636--536, with
$(\frac{F_{X,max}}{F_{X,pers}})=(\frac{F_{opt,max}}{F_{opt,pers}})^{\beta}$,
where $\beta$ varies with passband. Our burst shows that $\beta \sim 4$ which is comparable to the value $\beta_V \sim 3$ found for X1636--536
in the $V$ band, which is the closest approximation to our white light
passband.  Hence, we may imply that the
reprocessed emission from the GS\,1826--24 burst is also approximately that
from a blackbody (with a temperature set by the degree of X-ray irradiation), where the optical passband is on the Rayleigh-Jeans tail (Lawrence et al. 1983).

The X-ray burst observed by RXTE shows evidence for an increase in the 
burning area during the early rise phase, but no evidence for photospheric
radius expansion. Note that this is consistent with the fact that most
X-ray bursts showing photospheric radius expansion have rise times
less than $\sim$1 sec (see Lewin et al.\ 1995). However, by assuming that
our observed peak luminosity of 
$F_{max} = (2.8 \pm 0.4) \times 10^{-8}$ erg cm$^{-2}$ s$^{-1}$ is near
the Eddington limit of $\sim 1.8 \times 10^{38}$ ergs s$^{-1}$ for a 1.4
M$_{\odot}$ neutron star, we can set an upper limit to the distance to
GS\,1826$-$24. We derive a maximum
distance of $d = 7.5\pm0.5$ kpc. This estimate is consistent with the
upper limit from {\it
BeppoSAX} NFI observations ($7.4\pm0.7$ kpc; In't Zand et al. 1999) and the
optical lower limit of 4 kpc (Barret et al. 1995). The luminosity ratios
are  ${L}_{pers}/L_{burst}\sim$ 55 and $\sim$ 50 for the bursts observed
with {\it ASCA} and
{\it RXTE}, respectively. This is comparable with that found by {\it BeppoSAX} WFC  ($60\pm7$; Ubertini et
al. 1999). Coupling this value with an estimated stable
accretion rate of $\sim 1.5\times10^{-9}$ M$_{\odot}$ yr$^{-1}$ (Ubertini
et al. 1999), the burst must involve a combined hydrogen-helium burning
phase (Lewin, van Paradijs \& Taam 1995). This relatively long burst also
resembles the theoretical results of X-ray bursts driven by rapid proton
capture process, or rp-process (see Hanawa \& Fujimoto 1984; Taam 1981;
Bildsten 1998; Schatz et al. 1998).

Pedersen et al. (1982) have shown that the optical burst mainly reflects
the geometry of the system and that the contribution of
intrinsic radiative processes is small. Hence the correlated optical and
X-ray bursts discussed above are useful as probes of the structure and
geometry of the compact object surroundings.
Within the framework of the low-mass X-ray binary
system, the reprocessing can occur in the accretion disk and the companion
star.  Based on our observed mean delay of $3\pm1$ s for the
optical burst with respect to low energy X-rays,
we can then constrain the orbital period of the system. By Kepler's law,
the light travel time of 2--4 s corresponds to an orbital period of
1.6--5.5 hr if we assume a 1.4$M_{\odot}$ neutron star and a
companion star mass of 0.1--1.1$M_{\odot}$ (i.e. for a low-mass main sequence star and stable mass transfer). Hence, this range of periods
provides support for the $2.1\pm0.1$ hr orbital period proposed
by Homer et al. (1998). Lastly, from only one simultaneous optical/X-ray
burst, we cannot
draw a firm conclusion as to whether the optical burst is due to 
reprocessing in the disk or on the surface of the companion star.
However, given that the source is a low inclination system ($<$
70$^{\circ}$; Homer et al. 1998) and the ratio of smearing to delay is
$\sim 1$, the reprocessing is expected to be dominated by the accretion
disk. 

It is important in future studies to search for the possibly variable
delays if the reprocessing occurs on the surface of the companion star
(Matsuoka et al. 1984) or the `thick spot' in the disk proposed by
Pedersen et al. (1981). Whether the dominant reprocessor is the
companion star or the
`thick spot', one expects the ratio of optical to
X-ray flux in a burst to vary periodically. Moreover, the optical delay
would vary as a function of orbital phase as suggested by
Pedersen, van Paradijs \& Lewin (1981). Ubertini et al. (1999) recently
proposed a 5.76-hr
quasi-periodicity in the occurrence of X-ray bursts in GS\,1826--24, 
which makes this problem difficult to resolve with
current low-Earth-orbit satellites. However, with upcoming missions
such as {\it Chandra} and {\it XMM}, 
much longer continuous X-ray coverage will be possible, and together with
ground-based telescopes will enable us to probe the structure of this
source in much greater detail.

\section*{Acknowledgements}
We are grateful to Darragh O'Donoghue (SAAO) for his advice on the the use  
of the UCT-CCD for high-speed photometry and his help with the
subsequent reductions.  We also thank Lars Bildsten for valuable
comments and Fred Marang (SAAO) for his support
at the telescope, and the {\it RXTE} SOC team for their efforts in
scheduling the simultaneous time. This
paper also utilizes results provided by the ASM/{\it RXTE} team.
\bibliographystyle{../mn}

\end{document}